\documentclass[12pt]{article}
\usepackage{graphicx}
\usepackage{amsmath,amsfonts,amssymb,amsthm}
\usepackage{color}
\usepackage[a4paper=true]{hyperref}
\usepackage{enumerate}
\usepackage{color}

\theoremstyle{plain}
\newtheorem{theorem}{Theorem}

\theoremstyle{definition}
\newtheorem{definition}[theorem]{Definition}

\voffset = - 1.2cm
\title{A new Lie systems approach to second-order Riccati equations}

\author{J.F. Cari\~nena$^*$, J. de Lucas$^{**}$ and C. Sard\'on$^{***}$}

\begin{document}

\maketitle

\centerline{\footnotesize $^*$Departamento de F\'isica Te\'orica and IUMA, Facultad de Ciencias, Universidad de Zaragoza,}
\medskip
\centerline{\footnotesize Pedro Cerbuna 12, 50.009, Zaragoza, Spain.}
\medskip
\centerline{\footnotesize $^{**}$Institute of Mathematics, Polish Academy of Sciences,}
\medskip
\centerline{\footnotesize ul. \'Sniadeckich 8, P.O. Box 21, 00-956, Warszawa, Poland.}
\medskip
\centerline{\footnotesize $^{***}$\'Area de F\'isica Te\'orica, Facultad de Ciencias, Universidad de Salamanca,}
\medskip
\centerline{\footnotesize 37.008, Salamanca, Spain.}
\medskip

\begin{abstract}
This work presents a newly renovated approach to the analysis of second-order Riccati equations from the point of view of the theory of Lie systems. We show that these equations can be mapped into Lie systems through certain Legendre transforms. This result allows us to construct new superposition rules for studying second-order Riccati equations and to reduce their integration to solving (first-order) Riccati equations. 
\end{abstract}


\section{Introduction}	

A big deal of work has lately been devoted to the analysis of the so-called {\it Lie systems} \cite{LS}-\cite{Dissertations}. Such systems are characterized by admitting {\it superposition rules}, i.e. functions which determine their general solutions in terms of generic families of particular solutions and a set of constants. Lie systems also enjoy a plethora of geometric properties \cite{CGM00}. This makes them suitable for the investigation of many systems of differential equations and relevant problems in the physics, mathematics and control theory (see \cite{Dissertations} and references within).

The aim of this work is to present a new Lie systems approach to second-order Riccati equations. Such equations, whose interest relies on their appearance in physics and mathematics literature \cite{CRS05}-\cite{Er77}, had already been investigated from the viewpoint of Lie systems in several works \cite{CC87}-\cite{CL11Sec}. All these papers were based on the use of {\it ad hoc} changes of variables as well as some recent geometric techniques. However, we here introduce a further geometric method for the study of such equations, which is simpler and more powerful than previous ones. As a result, we derive a new superposition rule for analyzing second-order Riccati equations and prove that their explicit integration reduces to solving (first-order) Riccati equations \cite{Dissertations}. 
\section{Fundamentals on Lie systems}
For simplicity, we generally assume all functions to be real, smooth, and globally defined. In addition, we restrict ourselves to analyzing problems on linear spaces. In spite of this, our results can easily be generalized and proved in full detail through several minor technical modifications.

A {\it $t$-dependent vector field} over $\mathbb{R}^n$ is a map $X:(t,x)\in\mathbb{R}\times \mathbb{R}^n\mapsto X(t,x)\in {\rm T}\mathbb{R}^n$ 
such that $\tau\circ X=\pi_2$, where $\pi_2:(t,x)\in\mathbb{R}\times \mathbb{R}^n\mapsto x\in \mathbb{R}^n$ and $\tau:{\rm T}\mathbb{R}^n\rightarrow \mathbb{R}^n$ is the projection related to the tangent bundle ${\rm T}\mathbb{R}^n$. 

We call {\it integral curves} of a time-dependent vector field $X$ over $\mathbb{R}^n$ integral curves of the distribution spanned by its {\it suspension}, i.e. the vector field $\bar X=\partial/\partial t+X(t,x)$ over $\mathbb{R}\times \mathbb{R}^n$ \cite{FM}. Given an integral curve $\gamma:s\in\mathbb{R}\mapsto (t(s),x(s))\in \mathbb{R}\times \mathbb{R}^n$ for $X$ passing through the point $(t_0,x_0)\in\mathbb{R}\times \mathbb{R}^n$, there exists a time-reparametrization $\bar t=\bar t(s)$ 
such that
$$
\frac{d(\pi_2 \circ \gamma)}{d\bar t}(t)=(X\circ \gamma)(\bar t),
$$
which is referred to as the {\it associated system} related to $X$. Note that every $t$-dependent vector field determines a system of first-order differential equations of the above type and vice versa. 
This justifies to use $X$ to denote both a $t$-dependent vector field and its associated system.

\begin{definition} A {\it superposition rule} for a system of first-order differential equations 
\begin{equation}\label{Sys}
\frac{dx^i}{dt}=X^i(t,x),\qquad i=1,\ldots,n,
\end{equation}
is a function $\Phi:\mathbb{R}^{nm}\times \mathbb{R}^n\rightarrow
\mathbb{R}^n$, of the form $x=\Phi(x_{(1)}, \ldots,x_{(m)};k_1,\ldots,k_n),$ such that the general
 solution $x(t)$ of (\ref{Sys}) can be written as
  $x(t)=\Phi(x_{(1)}(t), \ldots,x_{(m)}(t);k_1,\ldots,k_n),$
where $x_{(1)}(t),\ldots,x_{(m)}(t)$ is any generic family of
particular solutions and $k_1\ldots,k_n$ is a family of constants to be related to initial conditions. 
 \end{definition}

Systems admitting a superposition rule are called Lie systems. The conditions guaranteeing that the system (\ref{Sys}) admits a superposition rule are stated by the {\it Lie--Scheffers Theorem} \cite[Theorem 44]{LS}. A modern statement of this relevant result is described
 next (we refer to \cite[Theorem 1]{CGM07} for details). 

\begin{theorem}{\bf (Lie--Scheffers Theorem)} A system (\ref{Sys}) admits a superposition rule if and only if the 
associated $t$-dependent vector field $X$ can be brought into the form $X_t=\sum_{\alpha=1}^rb_\alpha(t)X_\alpha$, 
for a certain family $b_1(t),\ldots,b_r(t)$  of $t$-dependent functions and a set of vector fields $X_1,\ldots,X_r$  spanning an $r$-dimensional Lie algebra, the so-called Vessiot--Guldberg Lie algebra.
\end{theorem}
 
The Lie--Scheffers Theorem enables us to reduce the integration  of a Lie system to solving a special type 
of Lie system on a Lie group. In fact, every Lie system $X$ on $\mathbb{R}^n$ associated with a Vessiot--Guldberg
 Lie algebra $V$, e.g. $X_t=\sum_{\alpha=1}^rb_\alpha(t)X_\alpha$ with $X_1\ldots,X_r$ a basis for $V$,
  can be associated with a (generally local) Lie group action $\Phi:G\times \mathbb{R}^n\rightarrow \mathbb{R}^n$ whose fundamental
   vector fields coincide with those of $V$. This action allows us to bring the general solution $x(t)$ of $X$ into the 
   form $x(t)=\Phi(g(t),x_0)$, where $x_0\in \mathbb{R}^n$ and $g(t)$ is the solution of the Lie system
\begin{equation}\label{EquLie}
\frac{dg}{dt}=-\sum_{\alpha=1}^rb_\alpha(t)X_\alpha^R(g),
\end{equation}
with $g(0)=e$ and $X^R_1,\ldots,X^R_r$ being a basis of right-invariant vector fields over $G$ closing on opposite commutation relations to $X_1,\ldots,X_r$ (see \cite{CGM00}). Conversely, the general solution of $X$ determines the solution
 for Eq. (\ref{EquLie}) \cite{AHW81,CarRamGra}.

Several methods can now be applied to solve Eq. (\ref{EquLie}) and, consequently, the Lie system (\ref{Sys}). If $\mathfrak{g}\simeq T_eG$ is solvable, the theory of reduction of Lie systems devised in \cite{CarRamGra} allows us to solve (\ref{EquLie}) by quadratures. More generally, we can use a  {\it Levi decomposition} of $\mathfrak{g}$ to write $\mathfrak{g}\simeq \mathfrak{r}\oplus_s \left(\mathfrak{s}_1\oplus\cdots\oplus\mathfrak{s}_l\right)$, 
where $\mathfrak{r}$ is the radical of $\mathfrak{g}$, $\mathfrak{s}_1\oplus\cdots\oplus\mathfrak{s}_l$ is the direct sum of a family 
of simple Lie subalgebras of $\mathfrak{g}$, and $\oplus_s$ denotes a semi-direct sum of $\mathfrak{r}$ and $\mathfrak{s_1}\oplus\cdots\oplus\mathfrak{s}_l$. 
The {\it theorem of reduction of Lie systems} \cite[Theorem 2]{CarRamGra} yields, with the aid of the previous decomposition, that the solution of a Lie system like  (\ref{EquLie}) but defined on a Lie group with 
Lie algebra $\mathfrak{s}_1\oplus\dots\oplus\mathfrak{s}_l$ enables us to construct a $t$-dependent change of variables that transforms Eq. (\ref{EquLie}) into a Lie system on a Lie group with 
Lie algebra $\mathfrak{r}$, which is integrable by quadratures  (see \cite{CarRamGra} for details). Summarizing, the explicit integration of (\ref{EquLie}) reduces to providing a particular solution of a Lie system related to $\mathfrak{s}_1\oplus\cdots\oplus\mathfrak{s}_l$. 

General solutions of Lie systems can also be investigated through superposition rules. There exist different procedures to derive them \cite{PW,CGM07,AHW81}, but we hereafter use the method
 devised in \cite{CGM07}, which is based on the {\it diagonal prolongation} notion \cite{CGM07,Dissertations}.

\begin{definition} Given a $t$-dependent vector field $X$ over $\mathbb{R}^n$, its  {\it diagonal prolongation} $\widetilde X$ to $\mathbb{R}^{n(m+1)}$ is the unique $t$-dependent vector field over $\mathbb{R}^{n(m+1)}$ such that:
\begin{itemize}
\item Given ${\rm pr}:(x_{(0)},\ldots,x_{(m)})\in \mathbb{R}^{n(m+1)}\rightarrow x_{(0)}\in \mathbb{R}^n$, ${\rm pr}_*\widetilde X_t=X_t$  $\forall t\in\mathbb{R}$. 
\item $\widetilde X$ is invariant under the permutations $x_{(i)}\leftrightarrow x_{(j)}$, with $i,j=0,\ldots,m$.
\end{itemize}
\end{definition}

The procedure to determine superposition rules described in \cite{CGM07} goes as follows. Take a basis $X_1,\ldots,X_r$ 
of a Vessiot--Guldberg Lie algebra $V$ associated with the Lie system. Choose the minimum integer $m$ so that 
the diagonal prolongations to $\mathbb{R}^{nm}$ of the elements of the previous basis are linearly independent at a generic point. 
Obtain $n$ functionally independent first-integrals $F_1,\ldots,F_n$ common to all the diagonal prolongations, $\widetilde X_1,\ldots,\widetilde X_r,$ to $\mathbb{R}^{n(m+1)}$, 
for instance, by the {\it method of characteristics}. Assume that these integrals take certain constant values, i.e., $F_i=k_i$ with $i=1,\ldots,n$. Employ these equalities to express the variables of one of the copies of $\mathbb{R}^n$ within $\mathbb{R}^{n(m+1)}$ in terms of the remaining variables and the constants $k_1,\ldots,k_n$. The obtained 
 expressions constitute a superposition rule in terms of any generic family of $m$ particular solutions and $n$ constants.

\section{A new Lie systems approach to second-order Riccati equations}

The most general class of second-order Riccati equations is given by the family of second-order 
differential equations of the form
\begin{equation}\label{NLe}
\frac{d^2x}{dt^2}+(f_0(t)+f_1(t)x)\frac{dx}{dt}+c_0(t)+c_1(t)x+c_2(t)x^2+c_3(t)x^3=0,
\end{equation}
with
$$
f_1(t)=3\sqrt{c_3(t)},\qquad f_0(t)=\frac{c_2(t)}{\sqrt{c_3(t)}}-\frac{1}{2c_3(t)}\frac{dc_3}{dt}(t), \qquad c_3(t)\neq 0.
$$
These equations arise by reducing third-order linear differential equations through a dilation symmetry and a time-reparametrization \cite{CRS05}. Their interest is due to their use in the study of several physical and mathematical problems
 \cite{CC87,GGL08,CL11Sec,CRS05,GL99}.

It was recently discovered that every second-order Riccati equation (\ref{NLe}) admits a $t$-dependent non-natural regular Lagrangian of the form
$$
L(t,x,v)=\frac{1}{v+U(t,x)},
$$
with $U(t,x)=a_0(t)+a_1(t)x+a_2(t)x^2$ and $a_0(t),a_1(t),a_2(t)$ being certain functions related to the
 $t$-dependent coefficients of (\ref{NLe}), see \cite{CRS05}. Therefore,
\begin{equation}\label{Legtrricsec}
p=\frac{\partial L}{\partial v}=\frac{-1}{(v+U(t,x))^2},
\end{equation}
and the image of the Legendre transform $\mathbb{F}L:(t,x,v)\in \mathcal{W}\subset \mathbb{R}\times {\rm T}\mathbb{R}\mapsto (t,x,p)\in \mathbb{R}\times {\rm T}^*\mathbb{R}$, where $\mathcal{W}=\{(t,x,v)\in \mathbb{R}\times {\rm T}\mathbb{R}\mid v+U(t,x)\neq 0\}$, is the open submanifold $\mathbb{R}\times\mathcal{O}$ where $\mathcal{O}=\{(x,p)\in {\rm T}^*\mathbb{R}\mid p< 0\}$. The Legendre transform is not injective, as $(t,x,p)=\mathbb{F}L(t,x,v)$ for $v={\pm 1}/{\sqrt{-p}}-U(t,x)$. Nevertheless, it can become so by restricting it to the open $\mathcal{W}_+=\{(t,x,v)\in \mathbb{R}\times {\rm T}\mathbb{R}\mid v+U(t,x)>0\}$. In such a case, $v=1/{\sqrt{-p}}-U(t,x)$ and we can define over $\mathbb{R}\times\mathcal{O}$ the $t$-dependent Hamiltonian
\begin{equation*}
 h(t,x,p)=p\left(\frac 1{\sqrt{-p}}-U(t,x)\right)-\sqrt{-p}=-2\sqrt{-p}- p\, U(t,x).
\end{equation*}
Its Hamilton equations read
\begin{equation}
\left\{
\begin{aligned}\label{Hamil}
\frac{dx}{dt}&=\frac{\partial h}{\partial p}=\frac{1}{\sqrt{-p}}-U(t,x)=\frac{1}{\sqrt{-p}}-a_0(t)-a_1(t)x-a_2(t)x^2,\\
\frac{dp}{dt}&=-\frac{\partial h}{\partial x}= p\frac{\partial U}{\partial x}(t,x)= p(a_1(t)+2a_2(t)x).
\end{aligned}\right.
\end{equation}

Since the general solution $x(t)$ of every second-order Riccati equation (\ref{Legtrricsec}) can be recovered from the general solution $(x(t),p(t))$ of its corresponding system (\ref{Hamil}), the analysis of the latter provides information about general solutions of second-order Riccati equations.

The important point now is that system (\ref{Hamil}) is a Lie system. Indeed, consider the vector fields over $\mathcal{O}$ of the form
$$
\begin{gathered}
X_1=\frac{1}{\sqrt{-p}}\frac{\partial}{\partial x},\quad\quad
X_2=\frac{\partial}{\partial x},\quad\quad
X_3=x\frac{\partial}{\partial x}-p\frac{\partial}{\partial p},\quad\quad
X_4=x^2\frac{\partial}{\partial x}-2xp\frac{\partial}{\partial p},\\
X_5=\frac{x}{\sqrt{-p}}\frac{\partial}{\partial x}+2\sqrt{-p}\frac{\partial}{\partial p}.\\
\end{gathered}
$$
Their non-vanishing commutation relations read
\begin{equation}\label{ComRel}
\begin{gathered}
\left[X_1,X_3\right]=\frac 12X_1,\qquad [X_1,X_4]=X_5,\qquad
[X_2,X_3]=X_2,\qquad [X_2,X_4]=2X_3,\\\left[X_2,X_5\right]=X_1,
\qquad \left[X_3,X_4\right]=X_4,\qquad [X_3,X_5]=\frac 12 X_5,\\
\end{gathered}
\end{equation}
and therefore span a five-dimensional Lie algebra $V$ of vector fields. Additionally, the $t$-dependent 
vector field $X_t$ associated with (\ref{Hamil}) holds
\begin{equation}
X_t=X_1-a_0(t)X_2-a_1(t)X_3-a_2(t)X_4.\label{F2}
\end{equation}
In view of expressions (\ref{ComRel}) and (\ref{F2}), system (\ref{Hamil}) is a Lie system. Note also that a similar result would have been obtained by restricting the Legendre transform over the open $\mathcal{W}_-=\{(t,x,v)\in\mathbb{R}\times{\rm T}\mathbb{R}\mid v+U(t,x)<0\}$.

Let us use the theory of reduction of Lie systems to reduce the integration of (\ref{Hamil}) to solving a Lie system on a Lie group. Using a Levi decomposition of $V$, we get $V\simeq V_1\oplus_sV_2$, with $V_2=\langle X_2,X_3,X_4\rangle$ being a semisimple Lie algebra isomorphic to 
   $\mathfrak{sl}(2,\mathbb{R})$ and $V_1=\langle X_1,X_5\rangle$ the radical of $V$. Hence, 
   $V$ is isomorphic to the Lie algebra of a Lie group $G=\mathbb{R}^2\rtimes SL(2,\mathbb{R})$, where
    $\rtimes$ denotes a semidirect product of $\mathbb{R}^2$ by $SL(2,\mathbb{R})$, and there exists a local 
    action $\Phi:G\times \mathcal{O}\rightarrow \mathcal{O}$ whose fundamental vector fields are 
     those of $V$. It is a long, but simple, computation to show that 
\begin{equation*}
\Phi\left(\left(
(\lambda_1,\lambda_5),
\left(\begin{array}{cc}
 \alpha&\beta\\
\gamma&\delta
\end{array}\right)\right),(x,p)\right)=\left(\frac{\sqrt{-\bar p}\bar x-\lambda_1}{\sqrt{-\bar p}+\lambda_5
},-(\sqrt{-\bar p}+\lambda_5)^2\right),
\end{equation*}
where $\bar x=(\alpha x+ \beta)/(\gamma x+\delta)$, $\bar p={p}\left(\gamma x+\delta\right)^2$ and $\alpha\delta-\beta\gamma=1$, is one of such actions (for a detailed example of how to derive these actions see \cite[Ch. 2]{Dissertations}).

The above action enables us to write the general solution $\xi(t)$ of system (\ref{Hamil}) in the form $\xi(t)=\Phi(g(t),\xi_0)$, 
where $\xi_0\in \mathcal{O}$ and $g(t)$ is the solution of the equation
\begin{equation}\label{HamGrup}
\frac{dg}{dt}=-\left(X^R_1(g)-a_0(t)X^R_2(g)-a_1(t)X^R_3(g)-a_2(t)X^R_4(g)\right),\qquad g(0)=e,
\end{equation}
on $G$, with the $X^R_\alpha$ being a family of right-invariant vector fields over $G$ whose vectors 
$X^R_\alpha(e)\in T_eG$ close on the same commutation relations as the $X_\alpha$ (cf. \cite{Dissertations}). 

We now turn to apply to Lie systems (\ref{HamGrup}) the theory of reduction for Lie systems. Since $T_eG\simeq \mathbb{R}^2\oplus_s
 \mathfrak{sl}(2,\mathbb{R})$, a particular solution of a Lie system of the form (\ref{HamGrup}) but over $SL(2,\mathbb{R})$, which amounts to integrating (first-order) Riccati equations (cf. \cite{AHW81,Dissertations}), provides us with a transformation which maps system (\ref{HamGrup}) into an easily integrable Lie system over $\mathbb{R}^2$. In short, the explicit determination of the general solution of a second-order Riccati equation reduces to solving Riccati equations. 

Another way of analyzing the solutions of (\ref{Hamil}) is based on the determination of a superposition rule. 
According to  the method sketched in Sec. 2, a superposition rule for a Lie system (\ref{Hamil}), which 
admits a decomposition of the form (\ref{F2}), can be obtained through two common functionally independent first-integrals for the 
diagonal prolongations $\widetilde{X}_{1},\widetilde{X}_{2},\widetilde{X}_{3},\widetilde{X}_{4},\widetilde{X}_{5}$ 
to a certain ${\rm T}^*\mathbb{R}^{(m+1)}$ provided their prolongations to ${\rm T}^*\mathbb{R}^{m}$ are linearly independent at a generic point. In our case,
  it can be easily verified that $m=4$. The resulting first-integrals, derived through a long but easy calculation (see \cite{CL11Sec} for a similar procedure), read
\begin{equation*}
\begin{aligned}
F_0=(x_{(2)}-x_{(3)})\sqrt{p_{(2)}p_{(3)}}+(x_{(3)}-x_{(1)})\sqrt{p_{(3)}p_{(1)}}+(x_{(1)}-x_{(2)})\sqrt{p_{(1)}p_{(2)}},\\
F_1=(x_{(1)}-x_{(2)})\sqrt{p_{(1)}p_{(2)}}+(x_{(2)}-x_{(0)})\sqrt{p_{(2)}p_{(0)}}+(x_{(0)}-x_{(1)})\sqrt{p_{(0)}p_{(1)}},\\
F_2=(x_{(1)}-x_{(3)})\sqrt{p_{(1)}p_{(3)}}+(x_{(3)}-x_{(0)})\sqrt{p_{(3)}p_{(0)}}+(x_{(0)}-x_{(1)})\sqrt{p_{(0)}p_{(1)}}.
\end{aligned}
\end{equation*}
Note that given a family of solutions $(x_{(i)}(t),p_{(i)}(t))$, with $i=0,\ldots,3$, of (\ref{Hamil}), then $d\bar F_j/dt=\widetilde X_tF_j=0$ for $j=0,1,2$ and $\bar F_j=F_j(x_{(0)}(t),p_{(0)}(t),\ldots,x_{(3)}(t),p_{(3)}(t))$.

In order to derive a superposition rule, we just need to obtain the value of $p_{(0)}$ from the equation $k_1=F_1$, where $k_1$ is a real constant, to get
$$
\sqrt{-p_{(0)}}=\frac{k_1+(x_{(2)}-x_{(1)})\sqrt{p_{(1)}p_{(2)}}}{(x_{(2)}-x_{(0)})\sqrt{-p_{(2)}}+(x_{(0)}-x_{(1)})\sqrt{-p_{(1)}}},
$$
and then plug this value into the equation $k_2=F_2$ to have
\begin{equation*}
\begin{aligned}
x_{(0)}&=\frac{k_1\Gamma(x_{(1)},p_{(1)},x_{(3)},p_{(3)})+k_2\Gamma(x_{(2)},p_{(2)},x_{(1)},p_{(1)})-F_0 x_{(1)} \sqrt{-p_{(1)}}}
{k_1(\sqrt{-p_{(1)}}-\sqrt{-p_{(3)}})+k_2(\sqrt{-p_{(2)}}-\sqrt{-p_{(1)}})-\sqrt{-p_{(1)}}F_0},\\
p_{(0)}&=-\left[{k_1/F_0(\sqrt{-p_{(3)}}-\sqrt{-p_{(1)}})+k_2/F_0(\sqrt{-p_{(1)}}-\sqrt{-p_{(2)}})+\sqrt{-p_{(1)}}}\right]^2,\\
\end{aligned}
\end{equation*}
where $\Gamma(x_{(i)},p_{(i)},x_{(j)},p_{(j)})=\sqrt{-p_{(i)}}x_{(i)}-\sqrt{-p_{(j)}}x_{(j)}$. The above expressions give us a superposition rule $\Phi:(x_{(1)},p_{(1)},x_{(2)},p_{(2)},x_{(3)},p_{(3)};k_1,k_2)\in{\rm T}^*\mathbb{R}^3\times\mathbb{R}^2\mapsto (x_{(0)},p_{(0)})\in {\rm T}^*\mathbb{R}$ for system (\ref{Hamil}). Finally, since every $x_{(i)}(t)$ is a particular solution for (\ref{NLe}), the map $\Upsilon=\tau\circ \Phi$ furnishes the general solution of second-order Riccati equations in terms of three generic particular solutions
 $x_{(1)}(t),x_{(2)}(t),x_{(3)}(t)$ of (\ref{NLe}), the corresponding $p_{(1)}(t),p_{(2)}(t),p_{(3)}(t)$ and two real constants
  $k_1,k_2$.
\section{Conclusions}
Our work introduces many advantages with respect to previous methods 
for studying second-order Riccati equations \cite{CC87}-\cite{CL11Sec}. Apart from avoiding the use of {\it ad hoc} changes of variables as in \cite{CC87,GGL08}, we directly transform second-order Riccati equations into Lie systems (\ref{Hamil}) by 
  Legendre transforms. This is much simpler than other approaches based on more elaborated geometric theories \cite{CL11Sec}, which map second-order Riccati equations into Lie systems associated with a $\mathfrak{sl}(3,\mathbb{R})$ Lie algebra. Indeed, our method reduces the explicit integration of second-order Riccati equations to solving Lie systems related to $\mathfrak{sl}(2,\mathbb{R})$, e.g. Riccati equations.
    
\section*{Acknowledgments}

J.F. Cari\~nena and J. de Lucas acknowledge partial financial support by research projects E24/1 (DGA), MTM2009-11154 and MTM2010-12116-E and C. Sard\'on acknowledges a fellowship from the University of Salamanca and  partial support by research project FIS2009-07880 (DGICYT).


\begin{thebibliography}{0}
\bibitem{FM}
R. Abraham, J.E. Marsden, 
{\it Foundations of Mechanics. Second Edition} (Addison--Wesley, Redwood City, 1987).

\bibitem{LS}
S. Lie and G. Scheffers,
{\it Vorlesungen \" uber continuierliche gruppen mit geometrischen und
  anderen Anwendungen}
(Teubner, Leipzig, 1893).

\bibitem{PW}
P. Winternitz,
Lie groups and solutions of nonlinear differential equations,  
{\it Lect. Notes Phys.} {\bf 189} (1983), 263--305.
%
\bibitem{CGM07}
J.F. Cari{\~n}ena, J. Grabowski and G. Marmo,
Superposition rules, Lie Theorem and partial differential equations,
{\it Rep. Math. Phys.} {\bf 60} (2007), 237--258.

\bibitem{CGM00}
J.F. Cari\~nena, J. Grabowski and G. Marmo,
{\it Lie--Scheffers systems: a geometric approach} (Bibliopolis, Naples, 2000).

\bibitem{CC87}
J. S. R. Chisholm and A. K. Common,
A class of second-order differential equations and related first-order systems,
{\it J. Phys. A:Math. Gen.} {\bf 20} (1987),  5459--5472.

\bibitem{GGL08}
I.A. Garc\'ia, J. Gin\'e and J. Llibre,
Li\'enard and Riccati differential equations related via Lie algebras,
{\it Discrete Contin. Dyn. S.} {\bf 10} (2008), 485--494.

\bibitem{CL11Sec}
J.F. Cari\~nena and J. de Lucas, 
Quasi-Lie schemes and second-order Riccati equations, {\it J. Geom. Mech.} {\bf 3} (2011), 1--22.

\bibitem{AHW81}
R. L. Anderson, J. Harnad and P. Winternitz,
Group theoretical approach to superposition rules for systems of Riccati equations,
{\it Lett. Math. Phys.} {\bf 5} (1981), 143--148.

\bibitem{CarRamGra}
J. F. Cari\~nena, J. Grabowski and  A. Ramos,
Reduction of time-dependent systems admitting a superposition principle,
{\it Acta Appl. Math.} {\bf 66} (2001), 67--87.

\bibitem{Dissertations}
J. F. Cari\~nena and J. de Lucas,
Lie systems: theory, generalisations, and applications, {\it Dissertationes Math.} {\bf 479} (2011), 1--162.

\bibitem{CRS05}
J. F. Cari{\~n}ena, M. F. Ra{\~n}ada and M. Santander,
Lagrangian formalism for nonlinear second-order Riccati systems: one-dimensional integrability and two-dimensional superintegrability,
{\it J. Math. Phys.} {\bf 46} (2005), 062703.

\bibitem{GL99} 
A. M. Grundland and D. Levi,
On higher-order Riccati equations as B\"acklund transformations,
{\it J. Phys. A:Math. Gen.} {\bf 32} (1999), 3931--3937.

\bibitem{Ar97}
C. Arnold,
Formal continued fractions solutions of the generalized second order Riccati equations, applications,
{\it Numer. Algorithms} {\bf 15} (1997), 111--134.

\bibitem{Er77}
L. Erbe,
Comparison theorems for second order Riccati equations with applications,
{\it SIAM J. Math. Anal.} {\bf 8} (1977), 1032--1037.

\end{thebibliography}
\end{document}